\documentclass{article}
\usepackage[utf8]{inputenc}

\title{Summary Report of AF1 to Snowmass 2021: Beam Physics and Accelerator Education within the Accelerator Frontier
}
\author{{\bf AF1 Conveners:} \\M.~Bai (SLAC), Z.~Huang (SLAC), S.M.~Lund (MSU/USPAS)}
\date{September 15, 2022}
\begin{document}

\maketitle

\section{Executive Summary}

Accelerator and beam physics (ABP) is the science of the motion, generation, acceleration, manipulation, prediction, observation and use of charged particle beams. The impressive advancements of accelerator frontiers is inseparable from fundamental accelerator and beam physics research and development. The quest of next generation accelerators and colliders for discovery science pushes accelerator science towards ultimate beams with unprecedented beam energy, intensity and brightness. Four grand challenges in beam intensity, beam quality, beam control and beam prediction were identified by the US Accelerator and Beam Physics (ABP) R\&D program~\cite{ABP}. AF community-wide discussion on the physics limits of ultimate beams also reached the conclusion that it is very difficult to reach the next generation of colliders that are another order of magnitude higher energy beyond proposed future colliders such as ILC, FCC and CEPC with point designs exploiting today's conventional accelerator technology. Hence, intensified R\&D in advanced accelerator concepts and technologies is needed to address these grand challenges and reach ultimate beams.

Nevertheless support for fundamental beam physics research has been declining. The NSF has narrowed its program in Accelerator Science, and funding by the DOE through the GARD and Accelerator Stewardship programs has been steady or declining. With the energy frontier shifted from the USA to Europe, maintaining needed collider expertise for concepts such as the $e^+e^-$ collider and muon collider has been difficult. This situation not only slows down advances, but also creates difficulty in maintaining the R\&D portfolio and retention of talent in the USA that will be necessary for future technical leadership. In addition, this declining support for accelerator research also threatens student training and work-force development in Accelerator Science and Engineering (AS\&E). 

To address these issues, we propose to further strengthen current US AS\&T R\&D efforts in directions that will create a robust and scientiﬁcally challenging R\&D program in accelerator and beam physics to position the ﬁeld of US High Energy Physics (HEP) to be productive and competitive for decades to come. This includes:
\begin{itemize}
    \item Establish a decadal road map of accelerator and beam physics research in the DOE OHEP General Accelerator R\&D (GARD) to address the four ABP "grand challenges".
    \item Re-establish a program of beam physics research on general collider-related topics to coherently guide efforts and exploit broad advances within the accelerator field that are presently less centered on future collider concepts. 
    \item Strengthen and expand capabilities of the US accelerator beam test facilities to maintain their competitiveness relative to worldwide capabilities. 
\end{itemize}

To build and maintain a strong diverse and inclusive workforce to support future HEP accelerator facilities, we also propose that the community pursue the following efforts to further strengthen Beam physics and Accelerator Science \& Engineering (AS\&E) education and outreach program:
\begin{itemize}
    \item Gather integrated statistics on workforce composition and needs as well as gender, and ethnicity for AS\&E students and workers in the labs, universities, and industry to monitor progress and better guide long-term efforts. This can be achieved by extending roles of the USPAS.
    \item The AS\&E field should organize a yearly undergraduate level recruiting program structured to draw in talent broadly and also enhance recruiting of women and underrepresented minority (URM) students. This could be coordinated with the USPAS.
    \item Increase US Particle Accelerator School (USPAS) office effort by one FTE to extend roles listed above to benefit the community; improve technical IT support of classes; and for long-range planning and stability.  
    \item Strengthen accelerator research at universities by funding professors and projects on campus and in collaborative lab efforts to increase visibility among undergraduates to recruit talent into the field.
    \item Lower the barriers to participation women and URM talent into the field and take concrete measures to improve discourse and support quality of life and family support issues at the labs to broadly retain talent. Recommended steps are detailed in the Education, Outreach, and Diversity section.   
\end{itemize}

\section{Education, Outreach, and Diversity}

We summarize the community-based consensus for improvements concerning education, public outreach, and inclusion in Accelerator Science and Engineering (AS\&E) that will enhance the workforce in the USA. The improvements identified reflect discussions held within the 2021-2022 Snowmass community planning process and are elaborated in Ref.~\cite{mei}. Although the Snowmass process centers on high-energy physics, Ref.~\cite{mei} outlines improvements for the full scope of U.S. AS\&E because HEP is now as likely, or even more likely, to benefit from advances driven by applications outside of HEP as it is for advances specifically targeted to future colliders for HEP. This present situation contrasts the the historic path of where accelerator advances for HEP colliders drove the broader field. It also is consistent with the DOE HEP acting as the steward of the full scope of AS\&T in the USA and the broad and active scope of ongoing accelerator projects in the USA. A skilled and productive AS\&E workforce can be drive needed advances and be drawn back into specific HEP collider projects as funding becomes available. Within this context, recommendations in Education, Outreach \& Diversity include:

\begin{itemize}
    \item {\bf Talent Recruitment:} Institute a yearly national undergrad-oriented recruiting class to draw in talent. This will benefit the DOE AS\&E Traineeships and the national labs to help reduce workforce shortages and help boost quality by drawing in the best talent. This might work best by: 
    \begin{itemize} 
      \item Implement within an expanded USPAS mission (see USPAS below). 
      \item Introduce the full breath of accelerator applications and highlight exciting and high impact technology at a survey level. This approach should complement present USPAS classes that are designed to boost the level of students already committed to the field rather than recruit talent into the field. 
      \item Formulate to draw in more women \& URM talent. 
      \item Fully fund students admitted, and based on interest, pair students with lab mentors/contacts for follow up to increase likelihood of the students remaining in AS\&E.  
    \end{itemize} 
    \item {\bf USPAS:} The US Particle Accelerator School functions primarily for graduate workforce training in AS\&E and is well functioning and broadly used within this context. Augmenting the school office by one FTE effort with appropriate skills will allow  improvement and augmentation for a variety of roles to benefit the community: 
    \begin{itemize} 
       \item Run national undergrad recruit class outlined above. 
       \item Regularly gather community statistics on jobs, workforce needs, diversity (ethnicity, gender, etc) to better guide workforce training needs for the DOE, the DOE Traineeships, national Labs, and support USPAS class selections. 
       \item Better support IT enhancements for education including: Cloud computing; Class videos and tutorials; Support use of interactive software tools for classes; Improved class web sites and shared materials to deliver more effective classes; Maintain and better disseminate course materials that are valuable to R\&D efforts. 
       \item Long-range planning to increase stability.
    \end{itemize} 
    \item {\bf Universities:} Universities are essential to drawing in workforce talent. When faculty projects are well supported, undergraduates in contact with faculty are drawn in to enhance recruiting and graduate students working under supported and connected faculty can be more effectively trained for high impact careers. The NSF-funded Center for Bright Beams\cite{CBB} (CBB) is an example of a high-impact Science and Technology Center that coordinates accelerator research and has been effective in drawing in talent while advancing project that improve a range of accelerator technology.  Funding for the CBB will terminate in four years. Previous NSF AS\&E funding calls have been put on hold and DOE Steweardship support is modest. This decay in support for university projects should be addressed by:
    \begin{itemize} 
      \item Increasing funding for research grants and programs to involve faculties and students in accelerator projects on both campus and in DOE lab facilities. Projects should be run with a longer range consistent with student timelines from recruitment to graduation.
      \item Supporting national lab programs and expectations to deliver colloquia at universities to facilitate connections with faculty and inspire and recruit student talent.  
    \end{itemize}
    \item {\bf DOE AS\&E Traineeships:} The national undergrad recruiting class described above can help draw in strong domestic talent needed in the Traineeships. Additionally, to better support present and future traineeships, the DOE should:
    \begin{itemize} 
      \item Set clear expectations on the DOE labs to support placement of traineeship students. 
      \item Relax severe cap limits on expenditures per student to better reflect actual graduate education costs and encourage broader university participation. 
      \item Allow international students to participate to preserve the strong historic benefits the USA has derived by drawing in the best talent possible.
    \end{itemize} 
    \item {\bf Diversity, Equity, and Inclusion:} 
    \begin{itemize} 
      \item To significantly augment the slow rise in diversity, the talent pool reached must be expanded. Promising approaches include: Enhance support to the national undergrad recruiting class described above to draw in women and URM talent; Supporting outreach colloquia at Minority Serving Institutes (MSIs).   
      \item Fund lab programs to: Address quality of life issues and family support such as child care; Improve the tone of professional discourse.
      \item Be more welcoming to underserved communities. Suggestions of approaches include: Strengthening connections to professional societies serving issues on diversity such as the  National Society of Black Physicists (NSBP) and the National Society of Hispanic Physicists (NSHP); Target more scholarship and fellowship support to draw in women and URMs (similar to examples in Ref.~\cite{urm-programs}); Parallel efforts on APS-IDEA teams at several national labs, universities, and institutions to stimulate suggestions for further improvement to the physics community; Encourage labs to reward employees for volunteering for outreach and inclusion efforts, making hirings addressing diversity, and improving community welcome.  
    \end{itemize}

\end{itemize}

\section{Accelerator and Beam Physics}
Beam physics has been a central part of modern accelerator science. A series of workshops to explore the direction and scope of the field were held by the US Accelerator and Beam Physics (ABP) R\&D program, primarily funded by the HEP GARD including the Accelerator Stewardship, as well as jointly by AF1, AF4 and AF6. Through these workshops, the grand challenges of beam physics and accelerator science emerged and physics limits of ultimate beams for future colliders were explored.

\subsection{Grand Challenges}
The US Accelerator and Beam Physics (ABP) R\&D program explores and develops the science of accelerators and beams to make future accelerators better, cheaper, safer, and more reliable. Particle accelerators can be used to better understand our universe and to aid in solving societal challenges~\cite{ABP}.

The primary scientific mission of ABP R\&D is to address and resolve the four Accelerator and Beam Physics Grand Challenges (GC): 
\begin{itemize} 
\item[] \textbf{Grand Challenge 1 (Beam Intensity):} How do we increase beam intensities by orders of magnitude?
\item[] \textbf{Grand Challenge 2 (Beam Quality):} How do we increase beam phase-space density by orders of magnitude, towards the quantum degeneracy limit?
\item[] \textbf{Grand Challenge 3 (Beam Control):} How do we measure and control the beam distribution down to the level of individual particles?
\item[]  \textbf{Grand Challenge 4 (Beam Prediction):} How do we develop predictive “virtual particle accelerators”?

\end{itemize} 

Other equally important ABP missions are associated with the overall HEP needs:
\begin{itemize}
    \item Advance the physics of accelerators and beams to enable future accelerators.
    \item Develop conventional and advanced accelerator concepts and tools to disrupt existing costly technology paradigms.
    \item Guide and help to fully exploit science at HEP accelerator R\&D beam facilities and operational accelerators.
    \item Educate and train future accelerator physicists.
\end{itemize}

We propose a robust and scientifically challenging program in accelerator and beam physics to address the Grand Challenges. This will help position the field of US HEP to be productive and competitive for decades to come. We also call for a systematic and organized effort in research into the early conceptual integration, optimization, and maturity evaluation of future advanced accelerator concepts. We emphasize that the accelerator and beam test facilities are critical to enabling groundbreaking research to address the Grand Challenges.  Finally, we emphasize that it is important to maintain consistent long-term support for existing cross-cutting educational mechanisms in the field of accelerator science and technology such as US Particle Accelerator School (USPAS), the Center for Bright Beams (CBB), and the DOE AS\&E Traineeships.  

\subsection{Research Areas}

Research community input during the two ABP workshops \cite{GARD1, GARD2, Summary} indicated the following areas of research are needed to address the above Grand Challenges (GC):
\begin{itemize} 
\item[] \textbf{Single-particle dynamics and nonlinear phenomena; polarized-beams dynamics.}
  \begin{itemize} 
    \item Impacts GC 1 and 2 and benefits from addressing GC 3 and 4.
  \end{itemize} 
\item[] \textbf{Collective effects (space-charge, beam-beam, and self-interaction via radiative fields, coherent synchrotron radiation, e.g.) and mitigation.}
  \begin{itemize}
    \item Impacts GC 1 and 2, and benefits from addressing GC 3 and 4.
  \end{itemize}
\item[] \textbf{Beam instabilities, control, and mitigation; short- and long-range wakefields.}
  \begin{itemize} 
    \item Impacts GC 1 and 2, and benefits from addressing GC 3 and 4.
  \end{itemize}
\item[] \textbf{High-brightness / low-emittance beam generation, and high peak-current, ultrashort bunches.}
  \begin{itemize} 
    \item Impacts GC 2, and benefits from addressing GC 3 and 4.
  \end{itemize} 
\item[] \textbf{Beam quality preservation and advanced beam manipulations; beam cooling and radiation effects in beam dynamics.}
  \begin{itemize} 
    \item Impacts GC 2, and benefits from addressing GC 3 and 4.
  \end{itemize} 
\item[] \textbf{Advanced accelerator instrumentation and controls.}
  \begin{itemize} 
    \item Impacts GC 3.
  \end{itemize} 
\item[] \textbf{High-performance computing algorithms, modeling and simulation tools.}
  \begin{itemize} 
    \item Impacts GC 4.
  \end{itemize} 
\item[] \textbf{Fundamental accelerator theory and applied math.}
  \begin{itemize} 
    \item Impacts all Grand Challenges.
  \end{itemize} 
\item[] \textbf{Machine learning and artificial intelligence.}
  \begin{itemize} 
    \item Impacts GC 3 and 4 in the short term and GC 1 and 2 in the long term.
  \end{itemize} 
\item[] \textbf{Early conceptual integration, optimization, and maturity evaluation of accelerator concepts.}
  \begin{itemize} 
    \item Focuses on science and technology gaps and bridges between the various R\&D efforts.
  \end{itemize} 
\end{itemize} 

The ABP research area shares many topics and physics issues with R\&D activities at other (non-HEP) labs and universities.

\subsection{Beam Test Facilities}

Demonstrating the viability of emerging accelerator and beam physics research ultimately relies on experimental validation. The US (both at national laboratories and universities), has a portfolio of beam test facilities capable of providing beams over a wide range of parameters that can be used to perform research critical to advancing AS\&T related to high-energy physics, basic energy science, and beyond. These accelerator test facilities have enabled groundbreaking  accelerator research essential to developing the next generation of energy-frontier and intensity-frontier user facilities; see Ref.~\cite{BTF} for an overview of the current portfolio.

The facilities include GARD-sponsored infrastructure whose principal mission is to support broad participation from the community of accelerator scientists including from Universities, Industry, and National Laboratories. These facilities enable research pertinent to APB and provide ideal platforms for training future accelerator scientists. They also provide ideal platforms to engage university faculty and their students in projects that can draw more talent into the field.  

There are several ABP research facilities, such as FACET (SLAC), AWA (ANL), ATF (BNL), BTF (ORNL), IOTA/FAST (FNAL) at the national labs, and CBETA (Cornell), MEDUSA (Cornell), PEGASUS (UCLA) and SAMURAI (UCLA) at universities. Such facilities are invaluable for advancing new accelerator concepts and technologies. But a significant fraction of them are aging, underfunded, or share infrastructure with user facilities which significantly reduces their potential to stimulate advances and draw in more talent. It is critical for the US accelerator program to provide robust funding to operate, maintain, and upgrade these accelerator test facilities so that they remain productive for ABP. Likewise, a green-field national facility should be considered to remain competitive with the significant infrastructure development starting in international facilities, e.g., in Europe~\cite{eupraxia, adolphsen2022european}. 


\subsection{Physics Limits of Ultimate Beams for Future Colliders}

The rapid development of accelerators and beams in the past century has led to incredible discoveries in physics, chemistry, biology, etc. To date, about 25 Nobel Prizes in Physics, and 7 in Chemistry, were enabled by significant contributions from improved accelerator and beams [1,3]. The quest of understanding matter and the fundamental forces of nature has pushed the energy of proton beams towards 10~TeV with high luminosity. Synchrotron light sources are pushing towards the diffraction limit, and the X-ray FEL is now reaching atto-second timescales. These developments have been transformative in fundamental physics, materials science, and biology.  

To support the next level of discovery in fundamental physics as well as other transformative science fields, particle beams with beyond the state-of-art performance are required. The hunt for new physics beyond our current understanding of the standard model pushes future colliders into an energy range beyond 10s of TeV. The discovery of new fundamental constituents with lepton colliders requires luminosity scaled with the center-of-mass energy $E_{\textrm{cm}}$  as $(\frac{E_{\textrm{cm}}}{10~\textrm{TeV}})^2\times 10^{35} \textrm{cm}^{-2}\;\textrm{s}^{-1}$. It is clear that to realise the ultimate colliders with the conventional RF-based accelerator technology requires either a leap-forward in developments of key technologies such as ultra high-field magnets and ultra high-gradient accelerating structures, or with significant increases in size and power consumption that results in staggering cost. Fig.~1 shows the maximum peak luminosity and size as a function of beam center-of-mass energy of colliders in the past, present, and those proposed for the future. Both future lepton and hadron colliders also have significant increases in size, both in terms of physical scale and economic cost.


It is evident that to reach ultimate collider energy and luminosity, dramatic advances in acceleration and beam technologies are required. Both laser-driven and beam-driven plasma wakefield acceleration, aka, LWFA and PWFA,  have been pursued and efforts have intensified worldwide. While unprecedented high acceleration gradients have been demonstrated with both PWFA and LWFA, the path towards TeV collider still requires numerous advances in physics and engineering to meet the repetition rate, staging requirements, and reliability to match or exceed performance that today's conventional accelerator technology has achieved. Nevertheless, as the advanced concept acceleration field rapidly advances, it is not yet appropriate to estimate associated performance limits at this point of time. The topics would clearly benefit from guidance by those experienced in the practical requirements of conventional colliders to keep advances better aligned with the long-term goal of achieving transformative parameters in a practical facility.      

\bibliography{}

\end{document}